
\magnification=1200
\baselineskip=18pt

\def\({\c c}
\def\|{\'\i }
\def\sqr#1#2{{\vcenter{\hrule height.#2pt
     \hbox{\vrule width.#2pt height#1pt \kern#1pt
      \vrule width.#2pt} \hrule height.#2pt}}}

\font\tituloa=cmb10 scaled\magstep1

\phantom{...}
\hfill UR-1369

\hfill ER 40685-819
\bigskip
\centerline{\tituloa A Note on Abelian Conversion of Constraints}
\vskip 2.0 cm
\centerline{Ricardo Amorim$\,^\ast$}
\centerline{ and}
\centerline{ Ashok Das$\,^\dagger$}
\bigskip
\centerline{\it Instituto de F\|sica}
\centerline{\it Universidade Federal do Rio de Janeiro}
\centerline{\it RJ 21945 - Caixa Postal 68528 - Brasil}
\vskip 1.5 cm
\centerline{\tituloa Abstract}
\bigskip

We show that for a system containing a set of general second class constraints
which are linear in the phase space variables, the Abelian conversion
can be obtained in a closed form and that the first class constraints
generate a generalized shift symmetry. We study in detail the example
of a general first order Lagrangian and show how the shift symmetry
noted in the context of BV quantization arises.

\vfill
\bigskip
\hbox to 3.5 cm {\hrulefill}\par
\item{($\ast$)} Electronic mail: ift01001 @ ufrj
\item{($\dagger$)} Permanent Adress: Department of Physics and Astronomy,
University of Rochester, Rochester, NY 14627, USA
\eject

\bigskip
In an attempt to unify the method of quantization for systems
containing both first class and second class constraints, Batalin, Fradkin
and Tyutin ( BFT ) [1] have proposed a systematic method for converting
all second class constraints in a theory to first class ones. The
idea of BFT, in simple terms, is to introduce additional variables
into the theory ( one for every second class constraint ) with a simple
Poisson bracket structure and transform the second class
constraints to a Taylor series in the new variables such that they
become first class. The simplest case, which has been studied in
detail [2]
( which also will be considered in this letter ) and where one assumes
that the new constraints are strongly involutive ( not just first
class ) is commonly referred as the Abelian conversion of the second class
constraints. The Hamiltonian of the system is, similarly, expressed
as a power series in the new variables and each term in the series is
determined by requiring that the new Hamiltonian is in involution
with the first class ( Abelian ) constraints. The original system is,
of course, obtained if one chooses a gauge condition where all the
new variables vanish. However, other gauge conditions may be more
useful depending on the system under study [2].

The Abelian conversion is an iterative procedure and, in general, the
new constraints and the Hamiltonian may not have a closed form. As a
result, even though the first class constraints correspond to generators
of symmetries [3] of the system, the question of symmetries
cannot be studied in general.
In this letter, therefore, we study a special class of second class
constraints and the symmetries generated by their Abelian conversion.
We show that for general second class constraints linear
in the phase space variables, the Abelian conversion of the
constraints and the Hamiltonian can be obtained in a closed form.
We identify the general form of the operator which transforms any
observable to its new form. The Abelian conversion, in this case,
leads to a generalized shift symmetry in such systems and the first
class constraints generate this symmetry. We show in detail, for a general
first order Lagrangian,  how the general results lead to
the shift symmetry [4] which has come to play an important role in the BV
quantization [5] of gauge theories and topological field theories. We would
like to point out here that a shift in the original variables arising
as a result of the Abelian conversion was already noted in the
context of a first order Lagrangian in ref. [6] and that was the
starting motivation for our detailed examination of the question of
symmetries.

Let us consider a Hamiltonian system with phase space variables
$y^\mu$, $\mu=1,2,...,2N$ and the canonical Hamiltonian $H_c(y)$.
We are considering here a system with a finite number of degrees of
freedom for simplicity only and the discussion generalizes  to
continuum theories in a straight forward manner. Let us assume that the
system has a set of second class constraints which are linear in the
variables $y^\mu$ and are denoted by

$$\chi_\alpha(y)\approx 0,\hskip .5cm \alpha = 1,2,...,2n,
\hskip .5cm n\leq N \eqno(1)$$

\noindent (There may be other constraints - first class and more complicated
second class constraints - in the system, but we will not be
concerned with them. ) By assumption [3], therefore,

$$\{\chi_\alpha,\chi_\beta\}=\Delta_{\alpha\beta}\,\eqno(2)$$

\noindent defines a constant, antisymmetric and invertible matrix.
 We can, of course,
write the Hamiltonian including the constraints as

$$H=H_c(y)+\lambda^\alpha\chi_\alpha\,,\eqno(3)$$

\noindent where $\lambda^\alpha$'s are the Lagrange multipliers which
can be determined by requiring the constraints in eq. (1) to remain invariant
under time evolution.

In order to convert the constraints in eq. (1) to first class
( Abelian ) ones, we introduce [1] additional variables $\psi^\alpha$,
$\alpha=1,2,...,2n$ and assume that the Poisson bracket structure

$$\{\psi^\alpha,\psi^\beta\}=\omega^{\alpha\beta}\eqno(4)$$

\noindent defines a constant, antisymmetric and invertible matrix. Given
this, we define new constraints

$$\tilde\chi_\alpha = \chi_\alpha + X_{\alpha\beta}\psi^\beta\eqno(5)$$

\noindent and require that

$$\{\tilde\chi_\alpha,\tilde\chi_\beta\}=0\eqno(6)$$

\noindent which leads to

$$X_{\alpha\beta}\omega^{\beta\gamma} X_{\delta\gamma}=
-\Delta_{\alpha\delta}\,,\eqno(7)\,.$$

\noindent We note from eqs. (2), (4) and (6) that $X_{\alpha\beta}$
can be chosen to be a constant, invertible matrix and the original
constraints can be converted to Abelian ones. Note that when $ \psi^\alpha
\approx 0$, the constraints in eq. (5) reduce to the original ones in
eq. (1).

We can transform the Hamiltonian following the method of BFT so that
it is in involution with $\tilde\chi_\alpha$. However, let us note
the following. Let

$$\eqalign {B_\alpha^\mu=&\omega_{\alpha\beta}
X^{\beta\gamma}\{\chi_\gamma,y^\mu\}\,\cr
\tilde y^\mu=&\psi^\alpha B_\alpha^\mu\cr}\eqno(8)$$

\noindent By definition, $B_\alpha^\mu$ is a constant matrix of rank $2n$
and we note that

$$\tilde H=H_c(y-\tilde y)\eqno(9)$$

\noindent is in involution with $\tilde \chi_\alpha$. In fact

$$\eqalign{\{\tilde\chi_\alpha,\tilde H\}=&\{\chi_\alpha,H_c(y-\tilde y)\}
+\{X_{\alpha\beta}\psi^\beta,H_c(y-\tilde y)\}\cr
=&{{\partial H_c(y-\tilde y)}\over{\partial y^\mu}}\{\chi_\alpha,y^\mu\}+
X_{\alpha\beta}{{\partial H_c(y-\tilde y)}\over{\partial
\tilde y^\mu}}\{\psi^\beta,\tilde y^\mu\}\cr
=&{{\partial H_c(y-\tilde y)}\over{\partial y^\mu}}(\{\chi_\alpha,y^\mu\}-
X_{\alpha\beta}B_\gamma^\mu\{\psi^\beta,\psi^\gamma\})\cr
=&{{\partial H_c(y-\tilde y)}\over{\partial y^\mu}}(\{\chi_\alpha,y^\mu\}-
X_{\alpha\beta}\omega^{\beta\gamma}B_\gamma^\mu)\cr
=&0\cr}\eqno(10)$$

It is needless to say that the Hamiltonian in eq. (9)
coincides with the transformed Hamiltonian
that will be obtained through the method of BFT.

Let us note that the operator

$$G= exp (-\tilde y^\mu{{\partial}\over{\partial y^\mu}})=exp (-\psi^\alpha
B_\alpha^\mu{{\partial}\over{\partial y^\mu}})\eqno(11)$$

\noindent acting on any observable transforms it into its new form.
In fact, it is straight forward to check that

$$\eqalign{G\,\chi_\alpha(y)=&\tilde\chi_\alpha=\chi_\alpha(y-\tilde y)\cr
G\,H_c(y)=&\tilde H= H_c(y-\tilde y)\cr
G\,A(y)=&\tilde A = A(y - \tilde y)\cr}\eqno(12)$$

\noindent Since the transformed observables ( including the
Hamiltonian ) depend only on $(y-\tilde y)$ for $2n$ such variables,
the system is invariant under the local shifts in these variables of
the form

$$\eqalign{y^\mu \rightarrow & y^\mu+ \epsilon^\alpha(t)X_{\alpha\beta}
\omega^{\beta\gamma}B_\gamma^\mu\cr
&= y^\mu+\alpha^\mu(t)\cr
\tilde y^\mu \rightarrow & \tilde y^\mu + \epsilon^\alpha(t)X_{\alpha\beta}
\omega^{\beta\gamma}B_\gamma^\mu\cr
&=\tilde y^\mu+\alpha^\mu(t)\cr}\eqno(13)$$

\noindent where $\epsilon^\alpha$ are $2n$ parameters of these local
transformations and it is straight forward to check that the
constraints $\tilde\chi_\alpha$ generate these shift symmetries.
We cannot, however, identify these symmetries yet with the shift
symmetries arising in the context of BV quantization [4] where there is a
shift invariance for every dynamical variable in the theory.

We end this general discussion with some interesting observations on such
systems. Note that if we define

$$z^\mu=y^\mu-\tilde y^\mu\eqno(14)$$

\noindent then

$$\{z^\mu,z^\nu\}=\{y^\mu,y^\nu\}+\{\tilde y^\mu,\tilde y^\nu\}\eqno(15)$$

\noindent Using the definitions in eqs. (8) and (7), we can then
show that

$$\eqalign{\{z^\mu,z^\nu\}=&\{y^\mu,y^\nu\} - \{y^\mu,\chi_\alpha\}
\Delta^{\alpha\beta}\{\chi_\beta, y^\nu\}\cr
=&\{{y^\mu,y^\nu\}}_D\cr}\eqno(16)$$

\noindent where the right hand side represents the Dirac bracket [3] of the
phase space variables  defined with respect to the second class
constraints in eq. (1). Next, we note that

$$\eqalign{\{\tilde\chi_\alpha,z^\mu\}=&\{\tilde\chi_\alpha,y^\mu\}
-\{\tilde\chi_\alpha,\tilde y^\mu\}\cr
=&\{\chi_\alpha,y^\mu\}
-X_{\alpha\beta}B_\gamma^\mu\{\psi^\beta,\psi^\gamma\}\cr
=&0\cr}\eqno(17)$$

\noindent which again follows from the definitions in eq. (8).
Physically, this simply means that since $\tilde\chi_\alpha$ generate
the same shifts in $y^\mu$ and $\tilde y^\mu$ (see eq.(13))
$(y^\mu-\tilde y^\mu)$ is invariant under the shifts. This also
implies that any observable with the functional form
$A(y^\mu -\tilde y^\mu)$ will be invariant under the shifts.

$$\{A(y^\mu -\tilde y^\mu),\tilde\chi_\alpha\}=0\eqno(18)$$

\noindent If we have another set of first class constraints in the theory,
namely,

$$A_i\approx 0\eqno(19)$$

\noindent then it is clear that under the Abelian conversion,

$$\{A_i(y^\mu -\tilde y^\mu),\tilde\chi_\alpha\}=0\eqno(20)$$

\noindent and that the structure of gauge algebra will remain invariant.

Let us next examine these results in the context of a general first
order Lagrangian of the form [7]

$$L={1\over 2}(p_i\dot x^i - x^i\dot p_i)-H_c(x,p)\eqno(21)$$

\noindent and show that the Abelian conversion generates the shift symmetry
crucial in the understanding of the BV quantization. The obvious
constraints in the first order formulation are

$$\eqalign{\chi_{1i}=&\pi^{(x)}_i-{1\over 2}p_i\approx 0\cr
\chi_2^i=&\pi^{(p)i}+{1\over 2}x^i\approx 0\cr }\eqno(22)$$

\noindent There may be other constraints present in the theory, but
we will concentrate only on these for our discussions. These
constraints can be easily checked to be second class, namely,

$$\{\chi_{1i},\chi_2^j\}=-\delta_i^j\eqno(23)$$

\noindent These are linear constraints in the phase space variables consistent
with our earlier discussion. However, it is worth noting that for
such a system, the number of second class constraints is the same as
the number of variables $(x^i,p_i)$ and second, by assumption the
Hamiltonian is independent of $\pi^{(x)}_i$  and $\pi^{(p)i}$ .

For the Abelian conversion, we introduce the additional variables
$(\tilde x^i,\tilde p_i)$ and assume that they satisfy the Poisson
bracket structure

$$\{\tilde x^i,\tilde p_j\}=\delta^i_j\eqno(24)$$

\noindent It is easy to determine now that

$$\eqalign{\tilde\chi_{1i}=&\pi^{(x)}_i-{1\over 2}p_i+\tilde p_i\approx 0\cr
\tilde\chi_2^i=&\pi^{(p)i}+{1\over 2}x^i-\tilde x^i\approx 0\cr }\eqno(25)$$

\noindent are the proper Abelian conversion of the constraints in eq.
(22), namely,

$$\{\tilde\chi_{1i},\tilde\chi_2^j\}=0\eqno(26)$$

\noindent It is also easily checked ( these can be derived
systematically through the method of BFT as well.) that

$$\tilde H=H_c(x-\tilde x,p - \tilde p)\eqno(27)$$

\noindent will be in involution with the first class (Abelian) constraints
$\tilde\chi_{1i}$ and $\tilde\chi_2^i$. For example,

$$\eqalign{\{\tilde\chi_{1i},\tilde H\}=&\{\tilde\pi^{(x)}_i,\tilde H\}+
\{\tilde p_i,\tilde H\}=-{{\partial\tilde H}\over{\partial x^i}}
-{{\partial\tilde H}\over{\partial\tilde x^i}} = 0\cr
\{\tilde\chi_2^i,\tilde H\}=&\{\tilde\pi^{(p)i},\tilde H\}+
\{\tilde x^i,\tilde H\}=-{{\partial\tilde H}\over{\partial p_i}}
-{{\partial\tilde H}\over{\partial\tilde p_i}} = 0\cr}\eqno(28)$$

\noindent In fact, we note that the operator

$$G=exp(-\tilde x^i{{\partial}\over{\partial x^i}}
-\tilde p_i{{\partial}\over{\partial p_i}}+
{1\over 2}\tilde p_i{{\partial}\over{\partial \pi^{(x)}_i}}
-{1\over 2}\tilde x^i{{\partial}\over{\partial \pi^{(p)i}}})\eqno(29)$$

\noindent acting on any observable transforms it to its new form. In
particular, we note that

$$\eqalign{G\chi_{1i}=&\tilde\chi_{1i}\cr
G\chi_2^i=&\tilde\chi_2^i\cr
G\,H_c=&\tilde H = H_c(x-\tilde x, p - \tilde p)\cr}\eqno(30)$$

It is worth noting here that the transformed Hamiltonian, $\tilde H$,
is invariant under the shift of all variables, namely,

$$\eqalign{x^i\longrightarrow &x^i +\epsilon_1^i(t)\cr
\tilde x^i\longrightarrow &\tilde x^i +\epsilon_1^i(t)\cr
p_i\longrightarrow &p_i +\epsilon_{2i}(t)\cr
\tilde p_i\longrightarrow &\tilde p_i +\epsilon_{2i}(t)\cr}\eqno(31)$$

\noindent However, the constraints $\tilde\chi_{1i}$ and $\tilde\chi_2^i$
are not invariant under these transformations. On the other hand, the
generalized shift transformations

$$\eqalign{x^i\longrightarrow &x^i +\epsilon_1^i(t)\cr
\tilde x^i\longrightarrow &\tilde x^i +\epsilon_1^i(t)\cr
\pi^{(p)i}\longrightarrow &\pi^{(p)i}+{1\over2}\epsilon_1^i(t)\cr}\eqno(32)$$

\noindent and

$$\eqalign {p_i\longrightarrow &p_i +\epsilon_{2i}(t)\cr
\tilde p_i\longrightarrow &\tilde p_i +\epsilon_{2i}(t)\cr
\pi^{(x)}_i\longrightarrow &\pi^{(x)}_i-{1\over2}\epsilon_{2i}(t)\cr}
\eqno(33)$$

\noindent leave the Hamiltonian, $\tilde H$ as well as the Abelian constraints
 $\tilde\chi_{1i}$ and $\tilde\chi_2^i$ invariant. Furthermore, it is
straight forward to check that  $\tilde\chi_{1i}$
generate the symmetry transformations in eq. (32) while $\tilde\chi_2^i$
generate those in eq. (33).

Let us finally note that the shift symmetry can be seen in the
Lagrangian formulation by noting that the on-shell Lagrangian ( where
the constraints hold ) has the form

$$\eqalign{\tilde L =&({1\over2}p_i-\tilde p_i)\dot x^i
-({1\over2}x^i-\tilde x^i)\dot p^i+{1\over2}\tilde p_i\dot{\tilde x^i}
-{1\over2}\tilde x^i\dot{\tilde p_i}- \tilde H\cr
=&{1\over2}(p_i-\tilde p_i)(\dot x^i-\dot{\tilde x}^i)-
{1\over2}(x^i-\tilde x^i)(\dot p_i-\dot{\tilde p}_i)-
H_c(x-\tilde x, p - \tilde p)\cr}\eqno(34)$$

\noindent where the shift symmetry is manifest. Written in the second
order formulation, this coincides with the Lagrangian considered in the
context of BV quantization [4]. However, it is worth pointing out here that
the Lagrangian in eq. (34) gives rise to four constraints which can
be written  in the equivalent form

$$\eqalign {\Phi_{1i}=&\Pi^{(x)}_i+\Pi^{(\tilde x)}_i\approx 0\cr
\Phi_2^i=&\Pi^{(p)i}+\Pi^{(\tilde p)i}\approx 0\cr
\Phi_{3i}=&\Pi^{(x)}_i-\Pi^{(\tilde x)}_i-(p_i-\tilde p_i)\approx 0\cr
\Phi_4^i=&\Pi^{(p)i}-\Pi^{(\tilde p)i}+(x^i-\tilde x^i)\approx 0\cr}\eqno(35)$$

The first two of these constraints are first class while the other
two are second class. This is what would be expected from the shift symmetry
discussed in the context of the BV quantization. However, in order
that our discussion does not give the impression that Abelianization
generates new constraints, we point out that the additional second
class constraints arise merely from writing a first order Lagrangian
for the tilde variables.

To summarize, we have shown that for a system of second class
constraints which are linear in the phase space variables, Abelian
conversion can be obtained in a closed form. We have identified the
operator which transforms any observable to its new form and have
shown that, in this case, the transformed theory has a shift symmetry
which is generated by the Abelian, first class constraints of the theory.
We have examined in detail an example of a general first order Lagrangian
and have shown how, in this case, the shift symmetries generated by
the Abelian conversion coincide with those studied in the context of
the BV quantization.
\bigskip

This work was supported in part by U.S. Department of Energy Grant
No. DE-FG-02-91ER40685 and by CNPq, Brazil.

\bigskip
{\tituloa References}
\bigskip
\noindent [1] I. A. Batalin and E. S. Fradkin, Nucl. Phys. {\bf B279}
 (1987)514;
I. A. Batalin and L. V. Tyutin, Int. Jour. Mod. Phys. {\bf A6} (1991)
3255 and references  cited therein. See also C. Wotzasek, Int. Jour.
Mod. Phys. {\bf A} (1990) 1123.

\noindent [2] T. Fujiwara, Y. Igarashi and J. Kubo, Nucl. Phys. {\bf B341}
(1990) 695; Y.Kim, S.Kim, W. Kim, Y. Park, K. Kim and Y.Kim, Phys. Rev.
{\bf D46} (1992) 4574;   R. Banerjee, Phys. Rev. {\bf D48} (1993) R54467;
R. Amorim and J. Barcelos-Neto, {\it BFT quantization
of the FJ chiral-boson }, to appear in Phys.Lett. {\bf B} (1994);
R. Amorim and J. Barecelos-Neto, {\it BFT quantization of
chiral-boson theories }, pre-print IF-UFRJ-20/94 , and references
cited therein.

\noindent [3] P.A.M. Dirac, Can. J. Math. 2 (1950) 129; {\it Lectures
on quantum mechanics} (Yeshiva University, New York, 1964);
A.Hanson, T.Regge and C.Teitelboim, Constrained Hamiltonian Systems,
Accad. Naz. dei Lincei (1976);
M. Henneaux and C. Teitelboim, Quantization of Gauge Systems,
Princeton Univ. Press (1992)

\noindent [4] J.Alfaro and P.H.Damgaard, Nucl. Phys. {\bf B}404 (1993) 751

\noindent [5] I.A. Batalin and G. A. Vilkovisky, Phys.Lett. 120{\bf B}
(1981) 27; I.A. Batalin and G. A. Vilkovisky, Phys.Rev. {\bf D28}
(1983) 2567 [E:{\bf D30}(1984) 508]; for a review, see F. De Jonghe,
PhD Thesis, Katholicke Universiteit Leuven (1993). See also M. Henneaux
and C. Teitelboim in ref. [3].

\noindent [6] R.Amorim, L.E.S.Souza and R.Thibes, BFT formalism and first
order systems, to appear in Z.Phys. {\bf C} (1994)

\noindent [7] R. Marnelius, Acta Phys. Pol. {\bf B}13 (1982)669;
R.Amorim and J. Barcelos-Neto, Z. Phys.{\bf C}38 (l988)643;

\bye